\documentclass{article}
\usepackage[
  colorlinks=true,
  linkcolor=blue,
  citecolor=red,
  urlcolor=blue,
  pdftitle={Perceptions of Agentic AI in Organizations: Implications for Responsible AI and ROI},
  pdfauthor={Lee Ackerman},
  pdfkeywords={Agentic AI, Responsible AI, Ethics, AI Implementation, Organizational Impact},
  bookmarksopen=true,
  bookmarksdepth=3
]{hyperref}
\usepackage{graphicx}
\usepackage{microtype}
\usepackage{ragged2e}
\usepackage[natbibapa]{apacite}
\usepackage{authblk}
\newcommand{\keywords}[1]{\textbf{Keywords:} #1}

\title{Perceptions of Agentic AI in Organizations: Implications for Responsible AI and ROI}
\author{Lee Ackerman\thanks{Email: Lee\_Ackerman@media-uni.de}}
\affil{MA AI and Societies, Media University of Applied Sciences}
\date{\today}

\begin{document}

\maketitle

\begin{abstract}
  As artificial intelligence (AI) systems rapidly gain autonomy, the need for robust responsible AI frameworks becomes paramount. This paper investigates how organizations perceive and adapt such frameworks amidst the emerging landscape of increasingly sophisticated agentic AI. Employing an interpretive qualitative approach, the study explores the lived experiences of AI professionals. Findings highlight that the inherent complexity of agentic AI systems and their responsible implementation, rooted in the intricate interconnectedness of responsible AI dimensions and the thematic framework (an analytical structure developed from the data), combined with the novelty of agentic AI, contribute to significant challenges in organizational adaptation, characterized by knowledge gaps, a limited emphasis on stakeholder engagement, and a strong focus on control. These factors, by hindering effective adaptation and implementation, ultimately compromise the potential for responsible AI and the realization of ROI.
\end{abstract}

\keywords{Agentic AI, Responsible AI, AI Ethics, Organizational Impact, Return on Investment (ROI), Organizational Perceptions, Interpretive Qualitative Research}

\section{Perceptions of Agentic AI in Organizations: Implications for
Responsible AI and
ROI}\label{perceptions-of-agentic-ai-in-organizations-implications-for-responsible-ai-and-roi}

As artificial intelligence (\textbf{AI}) systems rapidly gain autonomy,
ensuring their alignment with human values becomes critical \citep{Bostrom2018}. This paper explores how organizations are navigating the
complexities of agentic AI. Responsible AI frameworks, which guide the
ethical development and deployment of AI, include ethical guidelines,
transparency measures, accountability mechanisms, bias mitigation
strategies, privacy and data protection protocols, safety and security
standards, and stakeholder engagement processes. These frameworks are
informed by ethical principles (e.g., OECD AI Principles; \citealp{OECD2024})
and risk management guidelines (e.g., NIST AI Risk Management Framework;
\citealp{NIST2023}) which will
continue to evolve in response to the broader socio-technical
environment \citep{MacKenzie1999, Dignum2019, Floridi2023}.

This research adopts an interpretive qualitative approach to explore how
organizations perceive and adapt these frameworks in the context of
increasingly sophisticated agentic AI systems. The study focuses on
interpreting lived experiences of AI professionals who responded to an
online survey.

Agentic AI\footnote{\textbf{Generative AI Agents (Agentic AI)}:
  Generative AI agents, or Agentic AI, represent a more advanced
  category of AI agents that should be distinguished from simpler AI
  applications like AI assistants or chatbots. Generative AI agents
  leverage large language models (LLMs) and multimodal AI capabilities.
  These agents exhibit a higher degree of autonomy and adaptability,
  characterized by:

  \textbf{Emergent Behavior}: The ability to generate novel solutions,
  exhibit unexpected behaviors, and adapt to unforeseen challenges.

  \textbf{Multimodal Reasoning}: The capacity to process and integrate
  information from various sources, including text, images, audio, and
  video.

  \textbf{Proactive Planning}: The ability to autonomously plan and
  execute complex tasks, often involving multiple steps and interactions
  with the environment.

  \textbf{Continuous Learning}: The ability to continuously learn and
  adapt based on new information and experiences

  \textbf{Increased Organizational Demands}: Agentic AI systems often
  require more robust data infrastructure, advanced API integrations,
  and specialized organizational skills compared to simpler AI
  applications, raising the bar for successful implementation and
  responsible governance}, a new class of highly autonomous and
adaptable AI agents, leverages large language models (\textbf{LLMs}) and
multimodal AI capabilities to exhibit: emergent behavior, generating
novel solutions and adapting to unforeseen challenges; multimodal
reasoning, enabling them to process information from various sources
like text, images, and audio; proactive planning, giving them the
ability to autonomously plan and execute complex tasks; and continuous
learning, which allows them to adapt based on new information. The paper
investigates how organizations are adapting their responsible AI
frameworks to accommodate this novel technology and its unique
challenges and opportunities.

This context of rapidly evolving AI technologies leads us to the central
problem this research addresses: how do organizations\textquotesingle{}
perceptions of agentic AI influence their implementation of responsible
AI practices and their subsequent Return on Investment (\textbf{ROI})
calculations, including considerations of workforce skills and
capabilities?

\subsection{Contextualizing the Narrative: Literature
Review}\label{contextualizing-the-narrative-literature-review}

The adoption of generative AI has outpaced past technology launches like
the personal computer and the internet \citep{Bick2024}.
Our ethics -- the ideas of right and wrong along with supporting norms,
rules, and principles - become increasingly important given this growth
and potential for impact \citep{Pflanzer2022}.

Exploring AI and Ethics thoughtfully, requires considering AI's impact,
including its opportunities and risks \citep{Floridi2018}. However,
practitioners find ethical ideals abstract, open to interpretation, and
difficult to apply to AI, given AI's agency and the lack of
understanding of its inner workings \citep{Buijsman2025}. The challenges are further nuanced as AI performs work with
social dimensions -- cognitive work previously performed by humans --
inheriting the associated human responsibilities and prompting
discussions about advanced AI's potential \citep{Bostrom2014}.

Responsible AI Frameworks guide the development, deployment, and use of
AI systems to minimize potential harm \citep{Dehghani2024}. While
sharing common ethical principles like transparency, fairness,
responsibility, and privacy, practitioners are grappling with
differences in the details related to interpretation, importance, and
implementation \citep{Jobin2019}. Responsible AI Pattern
Catalogues highlight ongoing efforts to recognize proven solutions to
recurring problems, while also providing extensible and adaptable
structures for transitioning from principles to action \citep{Lu2024}. Legislation plays a crucial role in shaping responsible AI
practices. A key example is the EU Artificial Intelligence Act \citep{EU2024}, which takes a risk-based approach to regulating the
development and deployment of AI systems. However, ethical
considerations go beyond legal compliance. Despite long-standing
discussions on potential harms like biases, effective action has been
challenging, leading to technical debt \citep{Cunningham1992}, ethical debt
(as discussed in \citep{Field2024}), which refers to the accumulation of
ethical compromises, and governance debt \citep{Meskarian2023}, which
relates to the long-term consequences of neglecting governance
structures. Moving from theory to impact while also overcoming these
debts involves considerable work and investment, covering areas such as
organizational tactics, stakeholder management, and technical methods
\citep{Rakova2021}.

Generative AI, foundational to Agentic AI, has shortcomings in
reliability and learning, along with safety challenges and risks like
biases, privacy concerns, and over-reliance \citep{OpenAI2024, Bengio2025}. Cultural forces also shape hopes and fears, as people
balance the promise of ease with the fear of obsolescence \citep{Cave2019} and seek to overcome anxiety-inducing portrayals of AI \citep{Bo2024}. This interplay of novelty and familiarity can be
understood through Remediation \citep{Bolter2001}, where new media
refashions and repurposes older forms. Agentic AI remediates human
agency, communication, and automation, creating both excitement and
anxiety. Navigating this promise and uncertainty, requires understanding
the progression from AI Assistants to Agentic AI, with its advances in
autonomy, reasoning, adaptability, planning, and emergent behaviors
\citep{Thomas2024, NVIDIA2023, Russell2021, WEF2024, IBM2025}. Furthermore, examining Agentic AI's
emerging characteristics---autonomy, imperfections, motivations,
creativity---and consider its role as an ``actant'' (a participant in a
network of relationships) in complex interactions with the human and
digital world \citep{Kolt2025, Li2024} will raise questions about
workforce composition and the experience of human and digital workers
\citep{Biilmann2025}.

Agentic AI offers transformative opportunities for enterprises --
enabling unique work with precision and efficiency \citep{Bousetouane2025}.
This transformative value emerges as organizations deploy numerous
agents -- that are adaptable, intelligent, and domain specific -- to
support their needs \citep{McKinsey2024}. While ROI can be
measured through cost savings, revenue growth, or efficiency \citep{Chia2024}, justifying the investment in responsible Agentic AI requires a
broader view that includes the significant work, time, and resources
involved \citep{Bevilacqua2023}.

However, there is a crucial gap in understanding how practitioners
implement and adapt responsible AI frameworks when facing the unique
challenges of agentic AI. This research addresses this gap by examining
the perceptions and insights of those moving from theory to practice and
exploring the implications of agentic AI innovation for work, the
enterprise, and society. Specifically, it investigates how practitioners
perceive this environment, how it drives their organizational actions,
and how they measure the ROI of their responsible agentic AI
implementations.

\section{The Storytelling Process:
Methodology}\label{the-storytelling-process-methodology}

\subsection{Research Approach and
Rationale}\label{research-approach-and-rationale}

An interpretive qualitative approach was chosen to capture the rich
perspectives of industry professionals, prioritizing in-depth insights
over statistical generalizability. In this rapidly evolving field, these
practical applications offer valuable guidance and may inform future
quantitative studies.

This study utilized a concise survey via Microsoft Forms, with anonymous
data collection. A purposeful sampling approach targeted AI
professionals working with North American organizations (5,000+
employees or \$1B+ revenue), recruited through professional networks and
LinkedIn. The target sample size was 40-60 participants, prioritizing
thematic saturation and in-depth insights over statistical
generalizability.

\subsection{Ethical Considerations}\label{ethical-considerations}

Participants gave informed consent and were fully informed about the
study's purpose, time commitment, and data use. Anonymity was
maintained, no PII was collected, and data was stored securely. The
de-identified, aggregated dataset will be shared via GitHub, and
findings will be presented in aggregate form. The survey avoided biased
language, and efforts were made to ensure diverse participation. Key
terms, including agentic AI, were clearly defined in the survey and a
glossary. The research design and ethical considerations underwent
informal peer review.

\subsection{Study Limitations: Strengths and Weaknesses of the Chosen
Method}\label{study-limitations-strengths-and-weaknesses-of-the-chosen-method}

While valuable for capturing in-depth insights, the qualitative approach
has limitations. Recruitment from professional networks means the
research may not capture broader perspectives and lacks statistical
generalizability. Participants may have given responses reflecting
organizational policies rather than individual views. The dynamic nature
of AI and agentic AI's early stage may limit the insights long-term
value. Also, responsible AI practices around data protection, risk
mitigation, and privacy could have influenced responses in unaccountable
ways. However, the interpretive qualitative approach offers significant
strengths. It allows for an exploration that balances depth and breadth,
capturing nuanced experiences for a richer understanding of the topic.

\subsection{Research Tools and AI
Collaboration}\label{research-tools-and-ai-collaboration}

This research used a collaborative approach including a human
researcher, traditional methods, and generative AI tools. Literature
reviews were conducted using platforms like Google Scholar, Consensus,
arXiv, and EBSCO. AI tools--Gemini, ChatGPT, Copilot, and
Julius--assisted in design, analysis, drafting, and reviewing. Citations were managed using the apacite package in LaTeX. This
integrated approach demonstrated AI\textquotesingle s potential as a
research collaborator, enhancing efficiency and insight.

\subsection{Researcher Positionality and Bias
Mitigation}\label{researcher-positionality-and-bias-mitigation}

As the primary researcher, I oversaw the research design and
interpretation, using AI tools as collaborators. My background as a
technology practitioner and my studies in AI and Societies influenced my
approach. To mitigate biases in both my perspective and AI-generated
outputs, I used several strategies. AI-generated content was critically
reviewed against academic and industry sources, and fact-checking
addressed AI-generated inaccuracies and biases. Human oversight ensured
accurate representation of qualitative data, and AI collaboration
provided diverse perspectives to reduce personal bias.

\subsection{Ethical Research
Commitment}\label{ethical-research-commitment}

This research was conducted with a firm commitment to ethical practices
and scholarly integrity, with a focus on the ethical implications of AI
collaboration. Acknowledging my positionality, addressing AI biases,
maintaining a reflexive approach, and ensuring responsible AI
collaboration, I aimed to contribute responsibly, transparently, and
practically to the understanding of responsible AI in the age of agentic
AI.

\section{The Respondents' Stories:
Findings}\label{the-respondents-stories-findings}

\subsection{The Respondents}\label{the-respondents}

The study gathered insights from 44 professionals working with large
North American organizations (5,000+ employees or \$1B+ revenue). Most
respondents (approximately 60\%) were in the technology industry, with
other industries represented. Participants, including AI consultants,
developers, business leaders, and researchers, offered diverse
perspectives. Over 70\% of respondents had less than 5 years of
experience working with AI technologies (16 with 1-3 years, 11 with 3-5
years), reflecting the emergent nature of the agentic AI field. Their
collective experiences form the basis of this research\textquotesingle s
findings.

\subsection{Foundations for
Interpretation}\label{foundations-for-interpretation}

The domain of AI, particularly agentic AI, presents a multifaceted
challenge --- integrating technological advancements, socio-technical
considerations, and evolving organizational practices. Agentic AI
systems are inherently complex\footnote{It is important to distinguish
  between \emph{``complicated''} and ``\emph{complex''} systems. A
  complicated system, like a car engine, may have many parts, but its
  behavior is predictable and can be understood by analyzing its
  individual components. A complex system, like a rainforest or a
  multi-agent AI system, is characterized by interconnectedness,
  emergence, and unpredictability. In complex systems, the interactions
  between components are crucial, and the system\textquotesingle s
  behavior cannot be easily predicted or controlled by examining
  individual parts. The Cynefin framework \citep{Snowden2007}
  provides a useful model for understanding these differences and the
  appropriate approaches for managing them.}, even individually, and
this complexity amplifies in multi-agent systems, where organizations
face risks like miscoordination, conflict, collusion, manipulation, and
the propagation of errors, biases, and privacy loss, along with the
overriding of safeguards \citep{Hammond2025}. Understanding
individuals\textquotesingle{} lived experiences is essential, as their
narratives provide insights that are often missed in technical analyses.
This research explores the intricate dynamics of human interaction with
AI, recognizing these experiences' pivotal role in responsible
innovation.

Responsibility is paramount in ethical discussions. As \citet{Havel1990} stated, ``\ldots the only genuine backbone of all our actions --
if they are to be moral -- is responsibility. Responsibility is
something higher than my family, my country, my firm, my success.'' In
agentic AI, where ethical considerations are crucial,
Havel\textquotesingle s words compel us to prioritize responsibility and
morality, grounding our endeavors in ethical principles that transcend
individual or organizational interests. These narratives offer
contextualized insights that guide our understanding, ensuring that our
approaches to AI are based in the realities of human experience and
ethical considerations.

While acknowledging that these narratives offer a selective view of 44
respondents' experiences and that further exploration is needed, they
also provide contextualized insights that support future approaches to
AI in the realities of human experience and ethical considerations. This
research aims to inform and stimulate further investigation,
contributing to a more comprehensive understanding of
AI\textquotesingle s impact.

\subsection{Interpretive Synthesis}\label{interpretive-synthesis}

Figure 1's heatmap illustrates the interpretive synthesis framework,
with responsible AI\footnote{For this paper, Responsible AI is defined
  as designing, developing, and deploying AI systems in ethically,
  transparently, and alignment with societal values. It encompasses
  principles such as fairness, accountability, transparency, privacy,
  and inclusivity, aiming to minimize bias and harm while fostering
  trust. This definition and view of Responsible AI dimensions was
  informed by consulting the following sources:

  Google AI. (n.d.). AI Principles. Retrieved from
  \url{https://ai.google/responsibility/principles/}

  Gartner, Inc. (n.d.). Responsible AI. {[}Gartner Glossary{]}.
  Retrieved from
  \url{https://www.gartner.com/en/information-technology/glossary/responsible-ai}

  International Business Machines Corporation (IBM). (n.d.). Responsible
  AI. Retrieved from
  \url{https://www.ibm.com/think/topics/responsible-ai}

  Microsoft. (n.d.). Responsible AI. Retrieved from
  \url{https://learn.microsoft.com/en-us/azure/machine-learning/concept-responsible-ai}

  \url{https://openai.com/charter/}

  \url{https://openai.com/safety/}} dimensions on the Y axis and
emergent themes\footnote{The thematic framework, based on open-ended
  survey responses and refined using additional data, guided coding
  alongside Responsible AI dimensions. The resulting data was flattened
  and used to generate the heat map. Here's a quick summary of each
  dimension:

  Autonomy, Control, and Ethical Alignment: Balancing agent autonomy and
  human control to ensure ethical alignment.

  Organizational Culture, Practices, and Societal Impacts: Exploring the
  interplay between human-AI interaction and how agentic AI reshapes/is
  reshaped by organizations and society.

  The Strategic Importance of Responsible AI: Aligning agentic AI
  initiatives with strategy.

  Knowledge Gaps in the Emerging Agentic AI Landscape: Identifies and
  seeks to address knowledge gaps by building the necessary
  competencies.

  Challenges in Adapting Responsible AI Frameworks: Overcoming the
  challenges of adapting responsible AI frameworks to the rapid
  evolution of agentic AI.} on the X axis. The heatmap explores the
interconnectedness of themes and dimensions, a complex web influencing
responsible AI implementation. For instance, control desires intertwine
with knowledge gaps, which impact leadership, ethical debt, and
organizational change. The following sections explore each theme,
highlighting key interconnections and implications.

\begin{figure}
\centering
\includegraphics[width=0.99\textwidth]{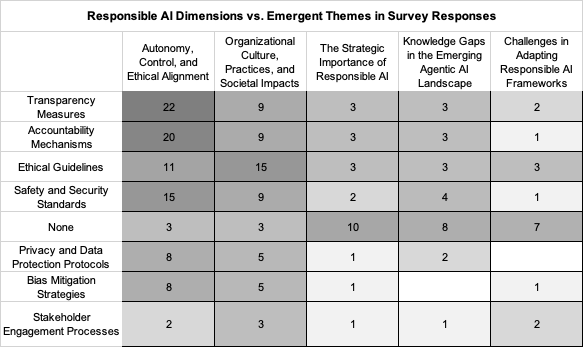}
\caption{Heat map showing how the dimensions for Responsible AI
intersect with the key themes.}
\end{figure}

\subsection{Autonomy, Control, and Ethical
Alignment}\label{autonomy-control-and-ethical-alignment}

Autonomy, control, and ethical alignment reveal a central tension in the
development and deployment of agentic AI: the desire to harness its
power while ensuring alignment with human values and oversight. One
respondent shared: ``Organizations must navigate regulatory uncertainty,
ensure transparency, and develop fail-safes to maintain control over
autonomous systems. How do you maintain safe, sustainable, scale?'' This
tension is echoed in concerns about ``control'', ``rules'',
``guidelines'', ``guardrails'', ``keeping humans in control'', ``kill
switches'', ``red-teaming'', ``fail-safes'', ``robust oversight'',
``ethical alignment'', and ``morality code integration.''

\begin{figure}
\centering
\includegraphics[width=0.99\textwidth]{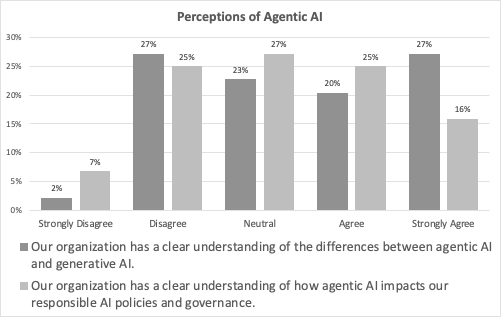}
\caption{Perceptions of Agentic AI in Organizations -
Distribution of Responses to Likert Questions}
\end{figure}

Organizations lack clarity on the differences between agentic AI and
generative AI (Figure 2), indicating a knowledge gap that impacts
responsible AI policies and governance. This is complicated by agentic
AI's recency \citep{NVIDIA2023} and challenges such as ``understanding AI
and uses,'' ``removing myths and fears,'' navigating the ``learning
curve,'' and even overcoming situations where ``vendors are putting
agentic labels on things that aren\textquotesingle t agents,'' as one
respondent noted. This uncertainty is echoed in Figure 3, where
open-ended responses often reflected generative AI concepts.

\begin{figure}
\centering
\includegraphics[width=0.7\textwidth]{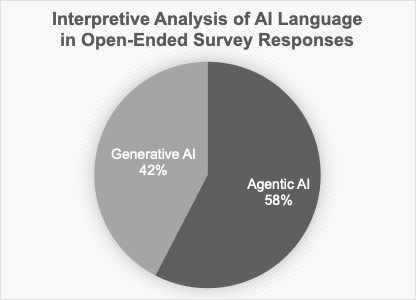}
\caption{This pie chart represents the results of an
interpretive analysis of AI-related language found in the open-ended
responses from the survey.}
\end{figure}

Complexity arises from layered knowledge gaps: first, the inscrutability
of the underlying LLMs themselves---even for their developers, as ``The
inner workings of these models are largely inscrutable, including to the
model developers'' \citep{Bengio2025}; and second, the knowledge
required to effectively build and deploy agentic AI on top of those
models. This combination of control desires and these fundamental
knowledge gaps is particularly salient given the potential for amplified
risks in multi-agent settings, such as miscoordination, conflict, and
bias propagation, which may be under-appreciated \citep{Hammond2025}.

Beyond this, a concern emerges -- are organizations sufficiently
equipped, with leadership, strategy, expertise, and the infrastructure,
to effectively create and operate responsible agentic AI? This leads to
a central question: How can we effectively respond? Is limiting autonomy
the answer? Or can we guide AI to be ethically aligned, even when it
operates with significant autonomy? Risk-averse may favour more control,
while those seeking first-mover advantage might accelerate the push to
autonomy.

\subsection{Organizational Culture, Practices, and Societal
Impacts}\label{organizational-culture-practices-and-societal-impacts}

\citet{Gruenert2015} observations on
organizational culture prompts defining tolerable behaviour in the age
of agentic AI. Respondents emphasized ``humans in the loop'' extends
beyond technical controls, impacting work and decision-making. Concerns
arose about AI outputs being used without reflection, highlighting the
need for clear frameworks and ethical guidelines. One participant shared
an optimistic view of the future: ``By recognizing and actively working
to reduce bias, we can harness the full potential of our data while
maintaining ethical standards and promoting fairness in our AI-driven
decisions.'' A culture prioritizing fairness and bias reduction must
invest in supporting practices.

Societal impacts extend beyond the workplace, raising concerns about
workforce replacement, privacy, and trust erosion. Respondents feared an
``overlord'' perspective, highlighting surveillance and control issues.
The delicate balance between automation and human agency was emphasized,
with concerns about AI replacing essential human decision-making.
Transparency and trust were identified as crucial, requiring user
experiences that avoid ``black box'' scenarios. One respondent shared:

\begin{quote}
The biggest challenge is making sure these AI systems are clear about
how they make decisions and that someone can be held responsible if
something goes wrong. Because agentic AI can make choices on its own,
it's sometimes hard to see exactly how it reaches those choices. That
makes it tough to fix mistakes or stop unfair behavior. Being open and
responsible about what AI does is really important to keep people's
trust.
\end{quote}

Controlling AI systems, including deactivation, raises questions about
responsibility and decision-making, especially when such decisions
benefit some users while harming others. As discussed in the previous
section on Autonomy, Control, and Ethical Alignment, implementing safety
mechanisms like red-teaming and fail-safes are crucial for maintaining
control and mitigating potential harms. However, beyond accountability,
proactive measures are needed to maintain trust. This includes securing
data, ensuring AI access doesn\textquotesingle t reveal or misuse
sensitive information, and implementing practices, audits, training, and
standard operating procedures. Bias mitigation strategies and
stakeholder engagement are crucial for ethical guidelines and control
mechanisms. To cultivate responsible AI, organizations must foster a
culture that prioritizes ethics, safety, security, privacy, inclusivity,
and accountability, and invest in supporting practices.

Ultimately, the data reveals human adaptation and ethical exploration.
Organizations are navigating cultural change and societal
responsibilities, not just deploying technology. Societal views of AI,
influenced by negative media portrayals, may shape respondents'
preconceptions \citep{Bo2024}. This re-presentation of older
technology anxieties in AI can be understood through
Remediation \citep{Bolter2001}, a tug of war where old and
new refashion each other. Agentic AI demonstrates this in work and
society. The challenge is fostering responsible innovation that
prioritizes human values and societal well-being.

\subsection{The Strategic Importance of Responsible
AI}\label{the-strategic-importance-of-responsible-ai}

Responsible AI\textquotesingle s strategic importance extends beyond
ethical compliance to critical business imperatives. Leadership
knowledge gaps, however, limit the strategic benefits. One respondent
noted, ``Information asymmetry among leaders in the organization. Before
they can lead change, they need to understand the value proposition of
AI agents, or even the basics of AI.'' This lack of understanding
hinders a strategic, systems-level approach to agentic AI, crucial given
its complexity and interconnected architectures. Leaders must understand
the technology, align the organization strategically, and manage a
workforce of digital and human workers \citep{Somers2025}.

\begin{figure}
\centering
\includegraphics[width=0.99\textwidth]{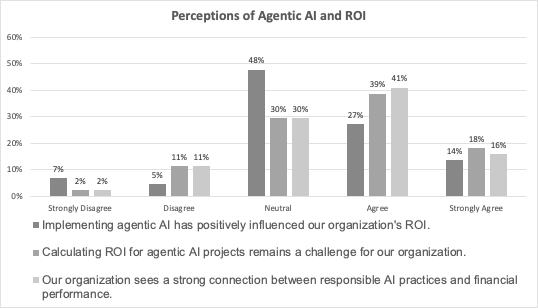}
\caption{Perceptions of Agentic AI's impact on financial
performance and ROI.}
\end{figure}

Figure 4 presents respondents\textquotesingle{} perceptions of agentic
AI and ROI, revealing potentially conflicting viewpoints. While 41\%
(27\% agree, 14\% strongly agree) indicate that implementing agentic AI
has positively influenced their organization\textquotesingle s ROI, a
substantial 60\% (7\% strongly disagree, 5\% disagree, 48\% neutral)
either disagree or are neutral on the subject. While neutral responses
are difficult to interpret, the positive responses are questionable as
we are so early in deploying agentic AI, there are knowledge gaps, and
for many organizations -- calculating ROI for agentic AI projects is a
challenge (only 13\% disagreeing that it is a challenge). Notably, a
substantial 57\% (41\% agree, 16\% strongly agree) see a strong
connection between responsible AI practices and financial performance.
Noting the confusion around generative AI vs. agentic AI, and the longer
amount of time organizations have had to work with generative AI, it may
be a case that respondents were answering this question based on
experience with generative AI. This data raises a critical strategic
concern: given leadership skill gaps, can organizations effectively
navigate this landscape, prioritize investments across initiatives and
time horizons, and ultimately realize agentic AI\textquotesingle s ROI?
How can leaders prioritize investments when they lack a fundamental
understanding of these initiatives?

Beyond these foundational challenges, liability and risk have emerged as
key strategic concerns. Neglecting responsible AI, particularly in areas
like bias and data security, can lead to significant financial risks,
legal liabilities, and reputational damage \citep{Bengio2025, Bevilacqua2023}. Such neglect leads to the accumulation of ``ethical debt'' -
the long-term consequences of ethical compromises - and ``governance
debt'' - the long-term consequences of neglecting governance structures
\citep{Field2024, Meskarian2023}. Illustrated by this
respondent\textquotesingle s comment:

\begin{quote}
Our company has accumulated decades of data, a valuable asset for
developing advanced AI systems. However, this extensive data may contain
inherent biases that can affect the performance and fairness of our AI
models. One of our biggest challenges with implementing AI systems is
addressing and mitigating these biases to ensure our AI solutions are
accurate, equitable, and reliable. By recognizing and actively working
to reduce bias, we can harness the full potential of our data while
maintaining ethical standards and promoting fairness in our AI-driven
decisions.
\end{quote}

Responsible AI is a strategic necessity, not just a guideline. Embracing
it fosters trust, enhances brand reputation, attracts and retains
talent, promotes innovation, ensures safety, and provides a competitive
advantage \citep{Bevilacqua2023}. Conversely, neglecting it risks
financial losses, legal battles, brand damage, and harm to communities
\citep{NIST2023, Kolt2025, Chan2023}. This strategic importance is echoed in several respondent
comments: ``I expect AI will be trusted more and more, and it will help
us find more efficiencies" and agentic AI will "enhance productivity
while keeping humans in control.'' It offers "global scale to agentic AI
and more monetization opportunities as companies, even individuals, will
have agents competing in a fabric-based marketplace, providing various
services.'' However, neglecting responsible AI hinders long-term
sustainability.

\subsection{Knowledge Gaps in the Emerging Agentic AI
Landscape}\label{knowledge-gaps-in-the-emerging-agentic-ai-landscape}

Agentic AI is in its nascent stage, with rapid evolution and constant
new information. This dynamic environment inevitably leads to knowledge
gaps, both for individuals and organizations. Acknowledging these gaps
is not a critique of the respondents\textquotesingle{} expertise, but a
recognition of the inherent challenges in this rapidly evolving domain.
Respondents described a ``hard to predict'' future, where change is
ongoing, and where ``\ldots we'll run into challenges first as we
learn.'' A respondent's simple statement of ``Don't know,'' when asked
about future trends highlights this exploratory moment.

Despite respondents\textquotesingle{} limited experience working with AI
technologies (Figure 5), knowledge gaps were surprisingly
underrepresented in the heat map (Figure 1). This inconsistency is
amplified by the data from Figure 2 which highlights perceptions that
organizations lack clarity on the differences between agentic AI and
generative AI -- and -- are unclear on how agentic AI impacts AI
policies and governance. This is further amplified by Figure 3 which
noted respondents potential knowledge gaps when it comes to the
differences between generative AI and agentic AI.

\begin{figure}
\centering
\includegraphics[width=0.8\textwidth]{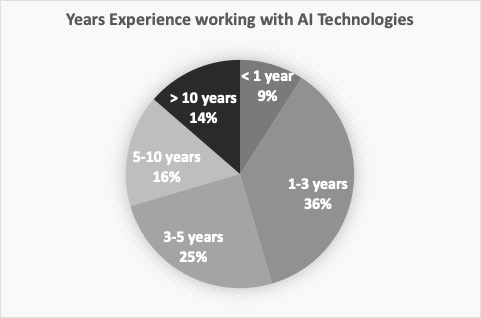}
\caption{Respondents reported years of experience working with
AI technologies}
\end{figure}

Agentic AI's dynamic nature compounds knowledge gaps, demanding
continuous learning. As it continues to evolve and grow in scale and
complexity, the knowledge gaps will widen if they are not proactively
addressed. As one respondent noted:

\begin{quote}
Right now we\textquotesingle re doing one-off agents and talking about
data and integration, which are essential but really table stakes.
Second big trend I think is that we will evolve communication frameworks
for agents to discover and communicate with each other. Think of it as
an extension of the agentic fabric architecture.
\end{quote}
 
\sloppy
These "table stakes", the ba\-sic, ear\-ly steps into the world of a\-gen\-tic
solutions, are already making the knowledge gap visible. These ideas of
agent fabrics represent massive advancements in complexity requiring
further growth in knowledge and an accelerated pace of learning. \citet{Spiegel2024} emphasizes the need for courage as we look ahead:
\fussy

\begin{quote}
It takes courage to imagine how future AI will ethically challenge our
conception of humanity and the world. And it takes courage to admit that
established ethical practices, beliefs, and theories are limited, and
therefore need not only be questioned, but also developed\ldots{}
\end{quote}

The ability to learn rapidly and continuously will be critical for
organizational success in agentic AI. As one respondent highlighted:

\begin{quote}
The most pressing challenge is to ensure that the workforce is ready for
the AI agentic era, i.e. people have the necessary skills to work with
agents, identify use cases for them, integrate them in their daily
workflows, and do all of this responsibly!
\end{quote}

Addressing knowledge gaps should be a top priority for leaders and be
reflected in organizational culture. Addressing these gaps through
targeted education, training, and dialogue is essential for fostering a
responsible and sustainable AI ecosystem, and ensuring the workforce is
prepared for the AI agentic era.

\subsection{Challenges in Adapting Responsible AI
Frameworks}\label{challenges-in-adapting-responsible-ai-frameworks}

Adapting responsible AI frameworks is impeded by agentic AI's inherent
ambiguity. One respondent noted: ``I think agentic systems will start as
glorified chatbots (many are, today, because vendors are putting agentic
labels on things that aren\textquotesingle t agents) and gradually gain
capabilities.'' This raises the question: Will experiences with
mislabeled and oversold capabilities frustrate those trying to support
and engage with change and adaptation?

\begin{figure}
\centering
\includegraphics[width=0.8\textwidth]{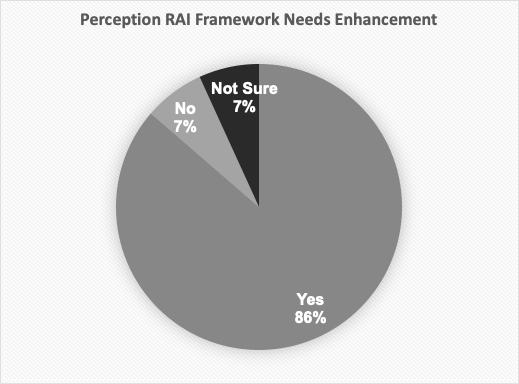}
\caption{A breakdown of responses to the question: "Do you
believe there is a need to enhance your organization\textquotesingle s
responsible AI framework to address the complexities of agentic AI?
(Select the best answer)"}
\end{figure}

Figure 6 shows a striking 86\% consensus: organizational responsible AI
frameworks need enhancement to address agentic AI complexities. This
clear mandate for change leads to the question: ``Which dimensions of
your responsible AI framework do you perceive as most likely needing
enhancement for agentic AI?''

\begin{figure}
\centering
\includegraphics[width=0.8\textwidth]{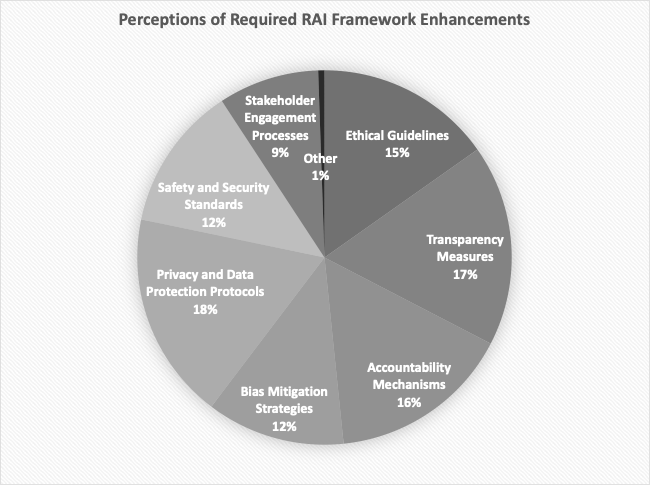}
\caption{Which dimensions of your responsible AI framework do
you perceive as most likely needing enhancement for agentic AI? (Select
all that apply)}
\end{figure}

As shown in Figure 7, ``Stakeholder Engagement Processes'' was the least
selected option. This is particularly concerning, as this topic was also
least discussed in the open-ended survey responses. Drawing from the
collaborative and creative practices of design leaders and design
thinkers, robust stakeholder engagement is crucial for solutions to
effectively meet user needs and achieve enhanced ROI \citep{Lockwood2018}. As \citet{Brown2019} notes, ``Complex systems have complex
stakeholders,'' and since agentic AI systems are complex, they too will
have complex stakeholders. It is also important to note that ``Your
ethical nightmares are partly informed by the industry you're in, the
particular kind of organization you are, and the kinds of relationships
you need to have with your clients, customers, and other stakeholders
for things to go well.'' \citep{Blackman2022}. The development of successful
and responsible agentic AI requires solutions that incorporate diverse
perspectives, actively support intended audiences, and empower those who
will maintain and operate these systems \citep{OECD2022, Floridi2018}. Furthermore, ``Stakeholder input is valuable, and responsible
decision-making involves it. But you cannot programmatically derive an
ethical decision just from stakeholder input. Whether you defer to or
defy (some subset of) stakeholder input, it's a qualitative ethical
decision.'' \citep{Blackman2022}.

Further research is needed to explore the low prioritization of
stakeholder engagement. This low prioritization may stem from
organizational culture, resource allocation, or knowledge gaps related
to agentic AI and engagement processes. Ideally, approaches like the
People + AI Guidebook \citep{GoogleAI2025} and Participatory AI \citep{Berditchevskaia2021}, emphasizing human-centered design and stakeholder co-creation,
gain traction. However, it\textquotesingle s crucial to recognize that
stakeholder engagement is just one piece of a larger ethical AI
strategy.

\begin{figure}
\centering
\includegraphics[width=0.99\textwidth]{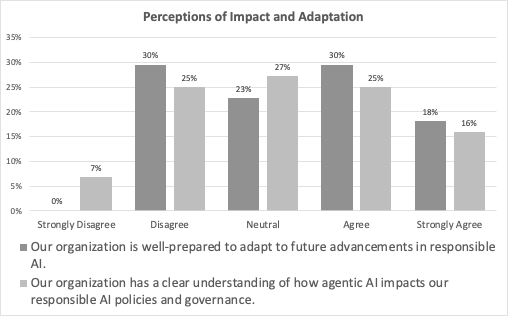}
\caption{Perceptions of organizations\textquotesingle{} ability
to change and understanding of impact}
\end{figure}

Figure 8 further underscores the adaptation challenges: only 48\% of
respondents felt ``well prepared for future advancements,'' and a
similar 41\% acknowledged a ``clear understanding of how agentic AI
impacts RAI policies and governance.'' This indicates a significant lack
of confidence and clarity among respondents regarding their
organizations' ability to adapt to the changing landscape. Combined with
the shared perception that there needs to be change and it is clear that
a gap between outcome and capability exists. This gap is significant -
as \citet{Blackman2022} highlights, responsible AI requires a comprehensive
approach encompassing AI ethical standards, organizational awareness,
dedicated teams and processes, expert oversight (such as an AI Ethics
Committee), accountability, an AI ethical risk program with KPIs, and
executive ownership. Other researchers have put forward the need for new
technical and legal infrastructure, supported by a governance strategy
based on principles of inclusivity, visibility, and liability \citep{Kolt2025}. Again, we see that our themes and dimensions are
interconnected---leadership, strategy, and practices converge on the
foundational need for robust stakeholder engagement.

\section{Methodological Considerations and Data Interpretation:
Capturing a Moment in
Time}\label{methodological-considerations-and-data-interpretation-capturing-a-moment-in-time}

This two-week online survey captured agentic AI perceptions from 44
North American professionals. While valuable, the geographically
dispersed sample and limited timeframe present inherent limitations. The
strong technology industry and AI consultant representation, coupled
with a respondent experience range of 1-5 years of working with AI
technologies, offers a snapshot of current, early experience
perspectives.

Data analysis revealed frequent confusion between generative and agentic
AI. To address this, responses were categorized using agentic
AI-specific terms like ``autonomous decision-making.'' However,
potential misinterpretations, overreporting, and social desirability
bias remained concerns across questions.

Future research should use clearer definitions and specific questions to
distinguish between anticipated and actual impacts. Participants could
define key constructs to reveal perspectives and knowledge gaps.
Qualitative methods, like interviews and focus groups, would offer
nuanced insights. Broader industry representation would validate
findings, identify biases, and introduce new perspectives.

\section{The Significance of the Stories:
Conclusion}\label{the-significance-of-the-stories-conclusion}

Agentic AI is complex and getting more complex as multi-agent solutions
scale. The challenge is magnified by an immature landscape that is
rapidly changing -- meaning that we're all still learning. While there
are gaps, challenges, and areas of friction -- there is also a clear
desire for responsible solutions -- and a healthy dose of caution.

Responsible AI Frameworks acknowledge the transformative potential of AI
systems, and guide the development, deployment, and use of such systems
in ways that minimize potential harm \citep{Dehghani2024}. But, as
highlighted in the introduction, practitioners grapple with
interpreting, prioritizing and implementing frameworks \citep{Jobin2019}. We need to close gaps. Gaps between theory and practice,
between potential and practice, between those designing and intended
users, and of course, the knowledge gap. The narratives shared serve as
a vital step towards fostering responsible and ethical agentic AI. While
this study captures a moment in this dynamic landscape, there are
numerous avenues for future exploration and studies.

For instance, organizations could take inspiration from the heat map
(Figure 2) to define a matrix of themes and responsible AI framework
dimensions that matter to their organization. Such a tool could help
them identify their gaps and then prioritize and plan their responsible
agentic AI efforts. To effectively navigate the tension between autonomy
and control in agentic AI systems, a concerted effort focused on
education and addressing knowledge gaps could be pursued. Noting the
incredible pace of change, learning should be lightweight, hands-on, and
strategically aligned. That is, the learning investment must be fully
aligned with the priorities of the organization. A focus should be
placed on dialogue and collaboration. Such an initiative could be
kickstarted by efforts around Participatory AI, include leaders, and be
tied to adjustments in organizational culture and practices. This path
forward should prioritize human-centric design, emphasizing meaningful
stakeholder engagement across development, business, and impacted
communities. Moreover, as agentic AI evolves, it\textquotesingle s
essential to critically examine its emerging characteristics---autonomy,
imperfections, creativity---and consider its role as an ``actant'' (a
participant in a network of relationships), with complex interactions
with our human and increasingly digital world \citep{Kolt2025, Li2024}. Seeing AI as more than just a ``tool'', is an important
adjustment for advancing its potential as part of the digital workforce,
and the potential of the entire workforce.

Through the combination of these efforts, a more nuanced approach to
control (and autonomy) will emerge as we better understand the
technology, failure modes, risk factors, and implications \citep{Hammond2025}. Leading us toward finding a responsible way to advance both
the human experience and the agent experience \citep{Biilmann2025}.

\section*{Acknowledgments}
I would like to express my sincere gratitude to the following individuals and institutions for their invaluable contributions to this research:
\begin{itemize}
    \item The faculty and staff of the MA AI and Societies program at Media University of Applied Sciences, for providing the academic foundation and support for this project.
    \item Professors Dr. Tong-Jin Smith and Dr. Jan-Henning Raff, for their guidance and feedback during the research design and development phases.
    \item Stefan Thissen, for his insightful discussions and peer review during the initial stages of this research, and my other classmates for their contributions to discussions during the initial stages of this research.
    \item My colleagues at TEKsystems Global Services, for sharing their expertise, providing feedback on the research design, and assisting in the recruitment of study participants.
    \item As part of this research, and reflecting the program's emphasis on AI and society, I utilized AI tools (Gemini, ChatGPT, Copilot, and Julius) to aid in research design, analysis, and drafting. This hands-on engagement allowed me to gain practical insights into the capabilities and limitations of these technologies.
    \item Finally, I thank the study participants for sharing their experiences, insights, and time.
\end{itemize}

\section{References}\label{references}
\bibliographystyle{apacite}
\bibliography{myreference}

\section{Appendix -- Additional Data
Visualizations}\label{appendix-additional-data-visualizations}

\begin{figure}[htbp]
\centering
\includegraphics[width=0.99\textwidth]{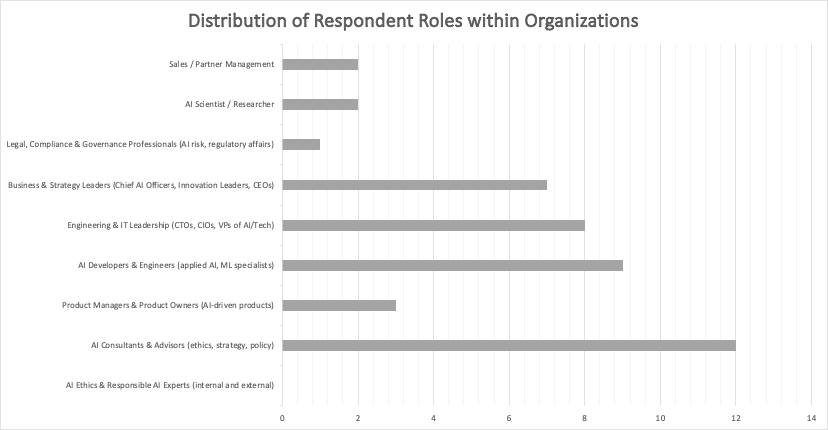}
\caption{Bar chart showing respondents answers to: What is your
role in your organization? (Select the best answer)". Note that some
categorization and grouping has been performed to answers provided to
the "Other" field.}
\end{figure}

\begin{figure}
\centering
\includegraphics[width=0.99\textwidth]{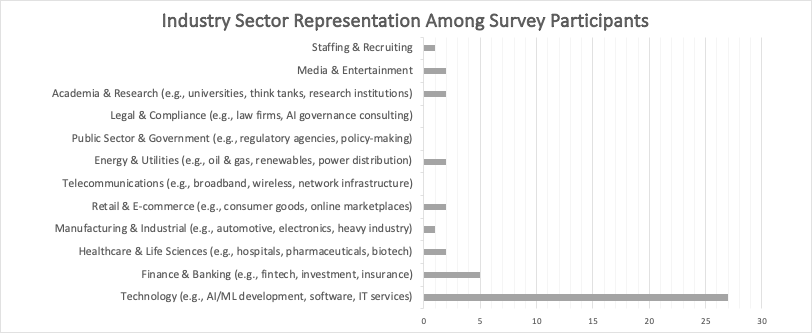}
\caption{Bar chart showing respondents answers to: "Which
industry best describes your organization\textquotesingle s primary
sector? (Select the best answer)". Note that some categorization and
grouping has been performed to answers provided to the "Other" field.}
\end{figure}

\begin{figure}
\centering
\includegraphics[width=0.99\textwidth]{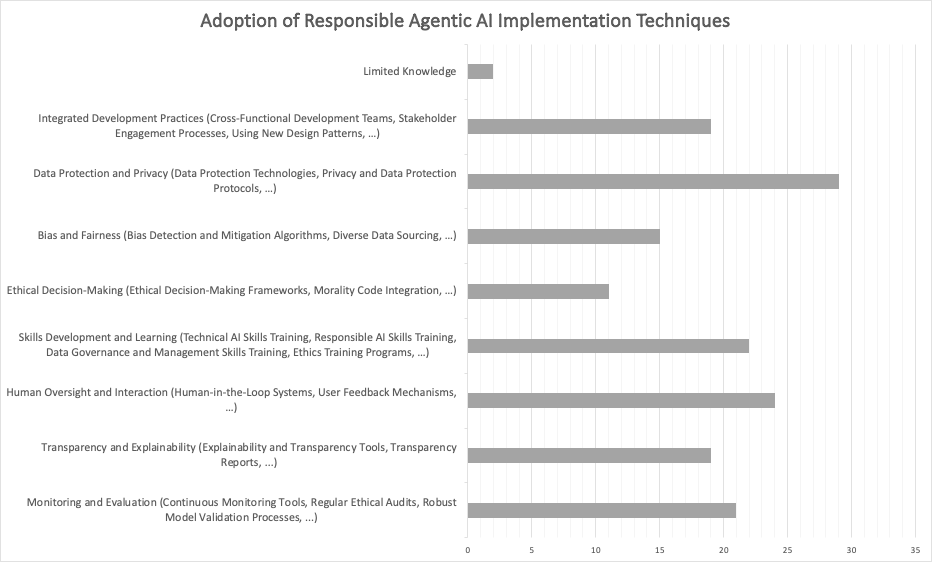}
\caption{Respondents answers to the question: "Which of the
following implementation techniques and technologies has your
organization adopted to ensure agentic AI operates responsibly? (Select
all that apply)". Note that some answers were recategorized from Other.}
\end{figure}

\begin{figure}
\centering
\includegraphics[width=0.99\textwidth]{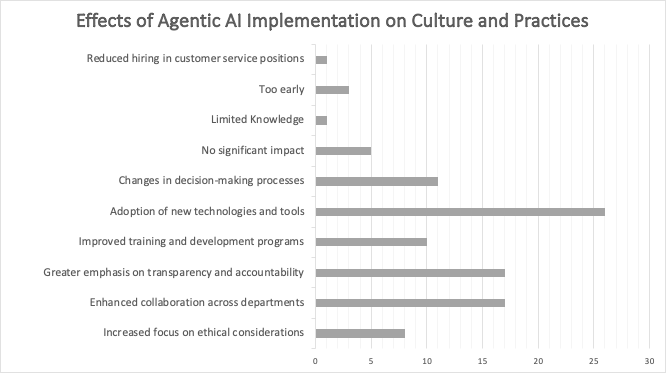}
\caption{Respondents answers to the question: "How has the
implementation of agentic AI influenced your
organization\textquotesingle s culture and practices? (Select all that
apply)". Note that some answers were recategorized from Other.}
\end{figure}

\begin{figure}
\centering
\includegraphics[width=0.99\textwidth]{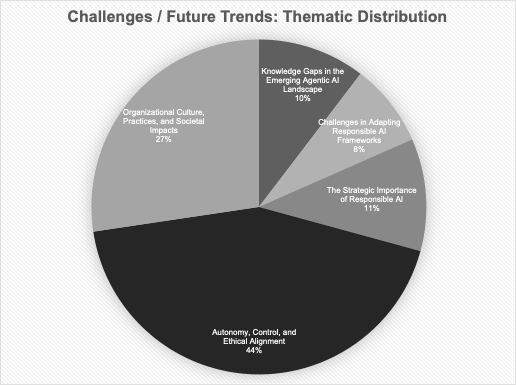}
\caption{Pie chart shows the distribution of themes across the
open-ended responses to the questions regarding current challenges and
future trends.}
\end{figure}

\begin{figure}
\centering
\includegraphics[width=0.99\textwidth]{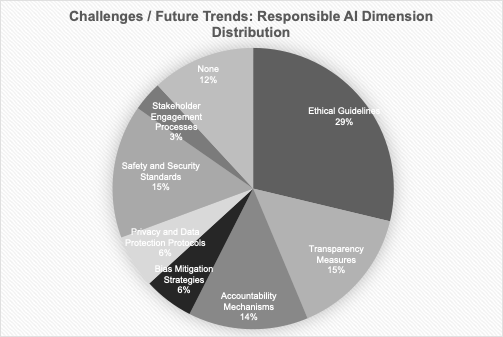}
\caption{Pie chart shows the distribution of responsible AI
dimensions across the open-ended responses to the questions regarding
challenges and future trends"}
\end{figure}

\begin{figure}
\centering
\includegraphics[width=0.99\textwidth]{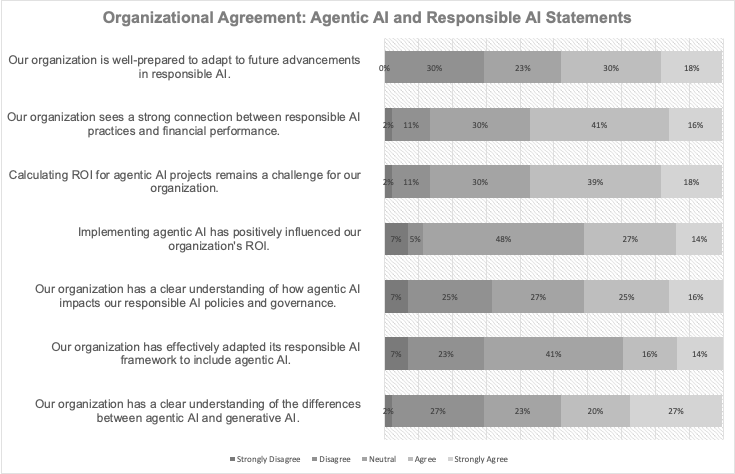}
\caption{Perceptions of Agentic AI in Organizations -
Distribution of Responses to Likert Questions}
\end{figure}
\end{document}